\documentclass[useAMS]{mn2e}
\usepackage{graphicx}
\usepackage{epsfig}
\usepackage{amsmath}
\usepackage{mathrsfs}
\usepackage{amssymb}
\usepackage[english]{babel}
\usepackage{amsfonts}
\usepackage{fontenc}
\usepackage{times}
\usepackage{txfonts}
\usepackage{textcomp}
\usepackage{booktabs}
\usepackage{microtype}
\usepackage{aas_macros}
\usepackage[pdftex]{lscape}
\usepackage[usenames,dvipsnames,svgnames,table]{xcolor}

\newcommand*{\Msun}{M_{\odot}}
\newcommand*{\Mstar}{M_*}
\newcommand*{\Minf}{M_{\rm inf}}
\newcommand*{\Msup}{M_{\rm sup}}
\newcommand*{\rssp}{r_{\rm SSP}}
\newcommand*{\Mto}{M_{\rm TO}}

\newcommand*{\Mej}{M_{\rm ej}}
\newcommand*{\Rssp}{R_{\rm SSP}}
\newcommand*{\Rira}{R_{\rm IRA}}
\newcommand*{\ai}{\alpha_i}
\newcommand*{\aij}{\alpha_{ij}}
\newcommand*{\bi}{\beta_i}
\newcommand*{\Iz}{I^{(0)}}
\newcommand*{\Izi}{\Iz_i}
\newcommand*{\Jz}{J^{(0)}}
\newcommand*{\Io}{I^{(1)}}
\newcommand*{\Jo}{J^{(1)}}

\newcommand*{\ci}{c_i}
\newcommand*{\yi}{y^{(i)}}
\newcommand*{\nei}{n_i}

\newcommand*{\Mtodot}{\dot{M}_{\rm TO}}

\newcommand*{\Nstep}{N_{\rm step}}

\title[Mass return from multiple stellar populations]
{A fast and accurate method to compute the mass return from multiple stellar populations} 

\author[Calura, Ciotti, \& Nipoti]  {F. Calura$^{1}$\thanks{E-mail:
     fcalura@oabo.inaf.it}, L. Ciotti$^{2}$, C. Nipoti$^{2}$ \\
(1) INAF, Bologna Astronomical Observatory, via Ranzani 1, 40127 Bologna, Italy\\ 
(2) Department of Physics and Astronomy, University of Bologna, viale Berti-Pichat 6/2, 40127 Bologna, Italy}

\date{Submitted, November 28, 2013}

\pagerange{\pageref{firstpage}--\pageref{lastpage}} \pubyear{2013}

\begin{document}
\maketitle
\label{firstpage}

\begin{abstract}

  The mass returned to the ambient medium by aging stellar populations
  over cosmological times sums up to a significant fraction (20\% -
  30\% or more) of their initial mass. This continuous mass injection
  plays a fundamental role in phenomena such as galaxy formation and
  evolution, fueling of supermassive black holes in 
  galaxies and the consequent (negative and positive) feedback
  phenomena, and the origin of multiple stellar populations in
  globular clusters. In numerical simulations the calculation of the
  mass return can be time consuming, since it requires at each time
  step the evaluation of a convolution integral over the whole star
  formation history, so the computational time increases quadratically
  with the number of time-steps. The situation can be especially
  critical in hydrodynamical simulations, where different grid points
  are characterized by different star formation histories, and the gas
  cooling and heating times are shorter by orders of magnitude than
  the characteristic stellar lifetimes.  In this paper we present a
  fast and accurate method to compute the mass return from stellar
  populations undergoing arbitrarily complicated star formation
  histories. At each time-step the mass return is calculated from its
  value at the previous time, and the star formation rate over the
  last time-step only. Therefore in the new scheme there is no need to
  store the whole star formation history, and the computational time
  increases linearly with the number of time-steps.

\end{abstract} 

\begin{keywords}
Galaxies: stellar content; galaxies: abundances; galaxies: ISM; methods: numerical
\end{keywords}

\section{Introduction}

For a Simple Stellar Population (SSP), the mass return rate from the
aging stars to the ambient medium depends on the relation between the
initial mass and the remnant mass of each star, and the Initial Mass
Function (hereafter IMF; e.g. Tinsley 1980, Matteucci \& Greggio 1986,
Tosi 1988, Ciotti et al. 1991, Maraston 1998). In general, the mass
return of a SSP represents a non-negligible fraction of its initial
mass, ranging from 20\% to 30\% for standard choices of the IMF (such
as a Scalo 1986, Chabrier 2003, Kroupa et al. 1993; e.g. Pellegrini 2012).

In stellar systems this source of fresh gas is present independently
of random phenomena such as galaxy merging; therefore, the mass return
of stellar populations plays a major role in determining the chemical
composition and the baryonic mass budget of the host systems.  For
example, the gas recycled by the aging stellar population is the main
mass source for gas flows in early-type galaxies such as cooling flows
and galactic winds (for general reviews see, e.g., Mathews \&
Brighenti 2003, Pellegrini 2012), for accretion of super-massive black
holes (SMBHs) at the center of spheroids (e.g. Norman \& Scoville
1988, Padovani \& Matteucci 1993, Tabor \& Binney 1993, Ciotti \&
Ostriker 1997, Shcherbakov et al. 2013, see also Ciotti \& Ostriker
2012 and references therein), and the consequent {\it negative and
  positive feedback} (e.g., Ciotti \& Ostriker 2007, Ishibashi \&
Fabian 2012, Zubovas et al. 2013).  Another case where the mass
returned from the evolving stars seems to play a fundamental role is
the origin of multiple stellar populations in globular clusters (see,
e.g., Piotto et al. 2007, Renzini 2008, D'Ercole et al. 2012).

Of course, real stellar systems are made by multiple stellar
populations, i.e. by a collection of SSPs assembled at different
epochs and with different metallicities, so that the mass return rate
is a function of their star formation history (SFH). In this case, an
accurate calculation of the mass loss rate requires keeping track of
the age and metallicity of each SSP and, depending on the desired time
resolution, its computation can be expensive in terms of computer
time. In fact, the function describing the mass return of a stellar
population is in general linked to its SFH through a convolution
integral over time. Therefore, not only the numerical evaluations
needed for its computation increase quadratically with the number of
time-steps, but also the whole SFH must be stored in the computer
memory. The situation becomes extremely time- and memory-consuming in
the case of hydrodynamical simulations, where each cell of the
numerical grid in principle hosts a different SFH, and where the
number of time steps can be of the order of millions or more for
simulations spanning a Hubble time, due to heating, cooling and
Courant times that can be orders of magnitude shorter than the
characteristic stellar lifetimes.

Similar (but less severe) problems affect also semi-analytic models
(SAMs) of galaxy formation (e.g., Baugh 2006).  In such models, the
evolution of the baryonic matter is driven by the evolution of the
dark matter halos, and the most massive galaxies form by the
progressive coalescence of a large number of progenitors. Thanks to
the simplified treatment of the various physical processes involved,
SAMs allow to explore the parameter space at a reasonable
computational cost.  However, in order to compute accurately the mass
return rate with an acceptable time resolution ($\simeq 10$ Myr or
less) over a Hubble time, the most massive galaxies require to store
the SFHs of several thousands progenitors (and this for an array
of $\simeq 10^2$ elements if one is interested in chemical evolution).
To reduce the computational cost, sometimes the mass return rate is
computed by means of the {\it Instantaneous Recycling Approximation}
(IRA), i.e. assuming that massive stars contribute instantaneously to
the mass return and to the chemical enrichment of the interstellar
medium (ISM), whereas the contribution of low and intermediate mass
stars is neglected at any epoch (e.g., Starkenburg et al. 2012).  This
approach has considerable limitations, since at short times after a burst of 
star formation 
it can lead to a severe overestimation of the
instantaneous mass return rate of the stellar populations
(e.g. Matteucci 2001). Note that the IRA is used also in
hydrodynamical simulations, for instance when studying the evolution
of star-forming molecular clouds (where the required time resolution
is set by the cooling time, of the order of $10^5$ yr, e.g. Krumholz
2011).

In order to overcome these problems and to compute accurately the mass
reprocessing from evolving stellar populations, in this paper we
present a fast and very accurate method which takes into account the
stellar lifetimes, at significantly reduced computational cost with
respect to the direct integration of the convolution integral. The
method bypasses the need of storing the SFH, as it uses information
relative to the previous time-step only, thus reducing the number of
evaluations of the convolution integral from quadratic to linear.  The
new scheme can be easily implemented in SAMs and in hydrodynamical
codes, with significant gain in accuracy and speed over standard
methods.  The basic idea extends a scheme already adopted in numerical
simulations to describe the delayed accretion of gas from the
accretion disk to the SMBH at the center of stellar spheroids, where
the mass flow on the accretion disk is determined by the solution of
the hydrodynamical equations for the ISM (e.g., Ciotti \& Ostriker
2007).  A simplified version of this method has been implemented in
Lusso \& Ciotti (2011) to describe the time evolution of the SNIa rate
in galaxy formation models with AGN feedback. In fact, the new method
here presented can be applied not only to the computation of the mass
return rate from stellar populations, but also to other astrophysical
studies where time-dependent convolution integrals with known kernels
must be computed (as in the case mentioned above of SNIa, e.g.,
Greggio 2005 for a full discussion).

The paper is organized as follows.  In Section 2 we introduce the
basic equations for the mass return rate and the formalism underlying
the new method. In Section 3 illustrative results are presented, and
finally in Section 4 we draw our conclusions. 
Mathematical detailes behind the method are given in the Appendix.

\section{The numerical problem} 

We start by considering a SSP of total mass $\Mstar$ and IMF described
by the function $\Psi(M)=\Mstar\phi(M)$; from now on all stellar
masses are in solar mass units, $\Msun$. The {\it normalized} IMF
$\phi(M)$ is defined so that
\begin{equation}
\int_{\Minf}^{\Msup} M \, \phi(M) \, dM = 1,
\label{eq_norm}
\end{equation}
where $\Minf$ and $\Msup$ are the minimum and the maximum stellar mass
in the population, respectively.

\begin{figure}
\epsfig{file=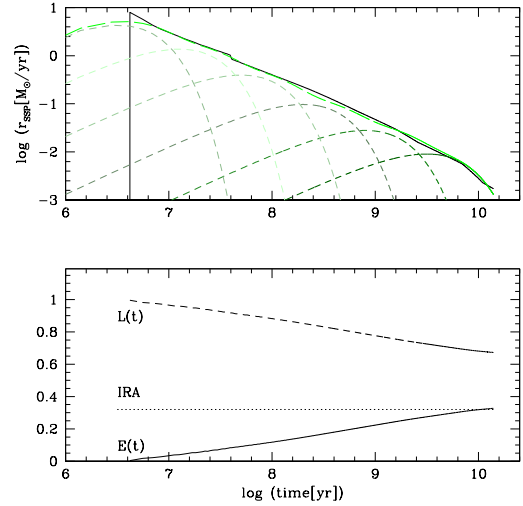,height=7.2cm,width=7.2cm}
\caption{Top panel: the exact, normalized $\rssp$ for a SSP of solar
  metallicity as given by eq. (2), is represented by the
  solid line.  The long-dashed line is the analytic fit obtained from
  eq. (11) with $k=6$, while the six components are represented by the
  short-dashed lines. The fit parameters are given in Table 1. Bottom
  panel: the exact cumulative mass fraction $E(t)$ (solid line,
  eq. 17), and the the locked-up fraction $L(t)=1-E(t)$. For
  reference, $E$ calculated in the IRA is represented by the dotted
  line.}
\label{rssp}
\end{figure}

As well known (e.g., Ciotti et al. 1991, Pellegrini 2012), for a SSP
assembled at $t=0$, the normalized mass return rate $\rssp$ can be
written as
\begin{equation}
\rssp (t)=\phi[\Mto (t)]\times \Mej [\Mto(t)]\times |\Mtodot (t)|,
\label{eqr}
\end{equation}
where $\Mto(t)$ is the mass of the stars entering the Turn-Off phase
at time $t$, and $\Mej(M)$ is the total mass ejected by a star of initial mass $M$
in the post main sequence evolutionary phases. By construction, the
mass return rate of the considered SSP is given by
$\Rssp=\Mstar\times\rssp$.

For a SFH characterized by a star formation rate $\psi(t)$, and a time
and metallicity independent IMF, the mass return rate at the time $t$
is given by
\begin{equation}
R(t) = \int_0^t \psi(\tau)\, \rssp (t-\tau) \, d\tau.
\label{rsfr}
\end{equation}
Note that the lower limit of integration is arbitrary, being related
to the beginning of star formation: for example, in this case we
assume without loss of generality that $\psi$=0 for $t<0$. On the
contrary, the upper limit of integration is physically required by the
obvious fact that mass return from a stellar population cannot take
place before its formation.

As pointed out in the Introduction, the integral above, when evaluated
by direct sum, scales quadratically with the number of time-steps
adopted to discretize the time $t$, so its computation can be
extremely time-consuming. We will now show how to evaluate eq. (3) as
a {\it linear} function of the number of time-steps, and without the
need of storing the whole evolution of $\psi (t)$. We will refer to
the rate as obtained by integrating eq. (3) with standard methods as
the {\it exact rate}, while the alternative scheme here presented will
be called the {\it new method}.

In the formalism of eq. (3), the mass return rate in the IRA is written as 
\begin{equation}
\Rira (t) =\psi(t)\int_0^{\infty}\rssp(\tau) d\tau :
\label{eq_ira}
\end{equation}
in practice, the IRA mass return rate is just given by the
instantaneous star formation rate times the total mass fraction
released by the SSP. Of course, the IRA overcomes all the problems of
computational times and memory storage posed by the standard
evaluation of eq. (3), but it can be properly used only when {\it all}
the timescales of the problem under consideration are much longer than
the lifetimes of the typical stars producing the bulk of the mass
return in the SSP.

\subsection{The new method} 

The idea behind the new method is to substitute the exact kernel
$\rssp$ in eq. (3) with a sum of functions with special mathematical
properties. We begin by illustrating the method for the case of a sum
of pure exponentials. Suppose that for a given SSP formed at $t=0$ the
associated $\rssp (t)$ can be represented with high accuracy as
\begin{equation}
\rssp (t) = \sum_{i=1}^k \ai e^{-\bi t}, 
\label{eq_fit0}
\end{equation}
where the parameters $\ai$ and $\bi$ are obtained by fitting the exact
function given in eq. (2); note that $\ai$ and $\bi$ are in units of
inverse of time (e.g., Gyr$^{-1}$).

It follows that eq. (3) can be written as
\begin{equation}
R (t) = \sum_{i=1}^k \ai\, \Izi (t), 
\label{eq_sum} 
\end{equation}
where 
\begin{equation}
\Izi (t)= \int_0^t\psi(\tau) e^{-\bi (t-\tau)}\,d\tau.
\label{eq_i}
\end{equation}
The meaning of the superscript ``(0)'' will become clear in the
following. For simplicity we drop the subscript index $i$,
as we now derive the generic expression for the family $\Iz$, and the
restoration of subscript $i$ in the resulting formulae is immediate.

It is straightforward to show that for a generic time interval
$\Delta t$ (not necessarily small), the function $\Iz$ can be
written rigorously as
\begin{equation}
\Iz (t+\Delta t) =  e^{-\beta\Delta t}\Iz (t)+\Jz (t,\Delta t), 
\end{equation}
where 
\begin{equation}
\Jz (t,\Delta t)\equiv\int_t^{t+\Delta t} \psi(\tau) \, e^{-\beta(t+\Delta t-\tau)}\,d\tau.
\end{equation}
In practice, at time $t+\Delta t$, each term of the sum in eq. (6) can be
calculated iteratively from the values at time $t$, plus
a contribution $\Jz$ due to the star formation over the last time
interval only.  For example, if we adopt a standard trapezoidal rule,
\begin{equation}
\Jz(t,\Delta t)\simeq {\Delta t\over 2}\left[\psi(t+\Delta t) + \psi(t)
   e^{-\beta\Delta t}\right ],
\end{equation}
and the final recursive formula for $R(t+\Delta t)$ is obtained by
summation of the $k$ components by using eq. (8).

Remarkably, for reasons that will be explained below, the method can
be extended by using in eq. (5) ``base'' functions more general than
pure exponentials, namely $t^n e^{-\beta t}$, where $n$ is an integer
(and, as discussed in the Appendix, not necessarily the same in all
the $k$ components).  In principle, this property allows to reproduce
quite complicated time dependencies of $\rssp(t)$, in particular the
initial rise due to the short but finite lifetimes of the most massive
stars in the IMF. For example, in the illustrative case presented in
Sect. 3 we found it optimal to use $n=1$ in all the
components. Accordingly, we now consider
\begin{equation}
\rssp (t) = \sum_{i=1}^k  \ai\, t \, e^{-\bi t}, 
\label{sum_rssp}
\end{equation}
(where of course $\ai$ and $\bi$ are different from those appearing in
eq. (5), and the $\ai$ are now in units of Gyr$^{-2}$). It follows
that 
\begin{equation}
R (t) = \sum_{i=1}^k \ai\, \Io (t), 
\label{eq_sum} 
\end{equation}
with
\begin{equation}
\Io(t)= \int_0^{t}\psi(\tau)\,(t-\tau)e^{-\beta (t-\tau)}\,d\tau.
\end{equation}
Simple algebra shows that for a generic time interval $\Delta t$, 
\begin{equation}
\Io (t+\Delta t)=e^{-\beta\Delta t}\left[\Io (t)+\Iz (t)\Delta t\right]+\Jo(t,\Delta t),
\end{equation}
where $\Iz (t)$ is still given by eqs. (7)-(8) with the new $\ai$
and $\bi$, and
\begin{equation}
\Jo(t,\Delta t)\equiv\int_t^{t+\Delta t}\psi(\tau)\,(t+\Delta
t-\tau)e^{-\beta (t+\Delta t-\tau)}\, d\tau.
\end{equation}
The integral $\Jo$ over the last time-step can be evaluated with a
simple trapezoidal rule, obtaining in this case
\begin{equation}
\Jo(t,\Delta t)\simeq {\Delta t^2\over 2}\psi(t) e^{-\beta\Delta t}.
\end{equation}

As anticipated above, the extension of the scheme of eqs. (8) and (14)
to functions $t^n e^{-\beta t}$ with integer $n\geq 0$ is
straightforward by using the binomial theorem, and the resulting
recursive formula for the functions $I^{(n)}(t+\Delta t)$ involves the
evaluation and the storing of the functions $I^{(j)}(t)$ for $j=0,1
,..., n$. One may ask why the method works, i.e., why it is possible
for this set of functions to evaluate a convolution integral over the
whole SFH just by keeping track (for each of them) of $n+1$ values
relative to the previous time. The reason is easily explained
following the general argument in the Appendix: here we just stress
that if the kernel $\rssp$ obeys a linear ordinary differential
equation (ODE) with constant coefficients, {\it then also $R$ obeys a
  linear ODE with constant coefficients of the same degree}, so that
integration of eq. (3) is equivalent to the integration of the
associated ODE, and the number of needed initial data is given by the
order of the equation.  For example, in the case of a single function
$t^n e^{-\beta t}$ the order of the ODE is $n+1$, the number of
quantities needed to be stored to evaluate $I^{(n)}$.  When $\rssp$ is
given by a sum of functions the argument is technically more
complicated, but conceptually identical, and following the Appendix
one can show that the specific case of eq. (11) requires the storing
of $2\,k$ values at the previous time.  For the more mathematically
inclined readers, in the Appendix we derive the ODE for $R(t)$ in the
case of the most general combination of functions $t^n e^{-\beta t}$
with arbitrary $n$, and we also briefly comment on the relation of the
present method with the theory of Green functions.

\section{Results}

In the example considered here for illustrative purposes of the
method, we evaluate $\rssp$ from eq. (2) by using $\Mej(M)$ as given
by Van den Hoeck \& Groenewegen (1997) for low and intermediate mass
stars ($0.8\le M\le 8$), and by Woosley \& Weaver (1995) for massive
stars ($M>8$).  In particular, the adopted $\Mej$ holds for a stellar
metallicity $Z=0.02$, very close to the concordance solar value of
0.015 (e.g, Lodders 2003).  For the IMF we adopt a Kroupa et al. (1993)
$\phi(M)$ with $\Minf=0.1$ and $\Msup = 40$, while the stellar
lifetimes are taken from Padovani \& Matteucci (1993).  Of course, the
general scheme of the method is independent of the specific choices
made, and different prescriptions for the properties of the SSP can be
considered as well (e.g., Ciotti et al. 1991, Pellegrini 2012). 

\subsection{Fitting the $\rssp$}
\label{single}

For the SSP described above Fig. 1 (top panel, solid line) shows the
exact normalized mass return rate $\rssp$ calculated according to
eq. (2) (top panel, solid line), and the corresponding returned
cumulative mass fraction as a function of cosmic time (bottom panel,
solid line)
\begin{equation}
E(t)=\int_{0}^{t}\rssp(\tau) d\tau. 
\end{equation}
The bottom panel also includes the quantity $L(t)=1-E(t)$, i.e. the
fraction of mass locked up in living stars and remnants (Portinari et
al. 2004). In this example, $E(\infty)\simeq 33$\% and
$L(\infty)\simeq 67$\%. For comparison, the dotted line represents the
cumulative returned mass fraction calculated assuming the IRA, and it
is apparent how this assumption leads to overestimate the mass return
of a SSP, in particular at early times.

\begin{figure*}
\epsfig{file=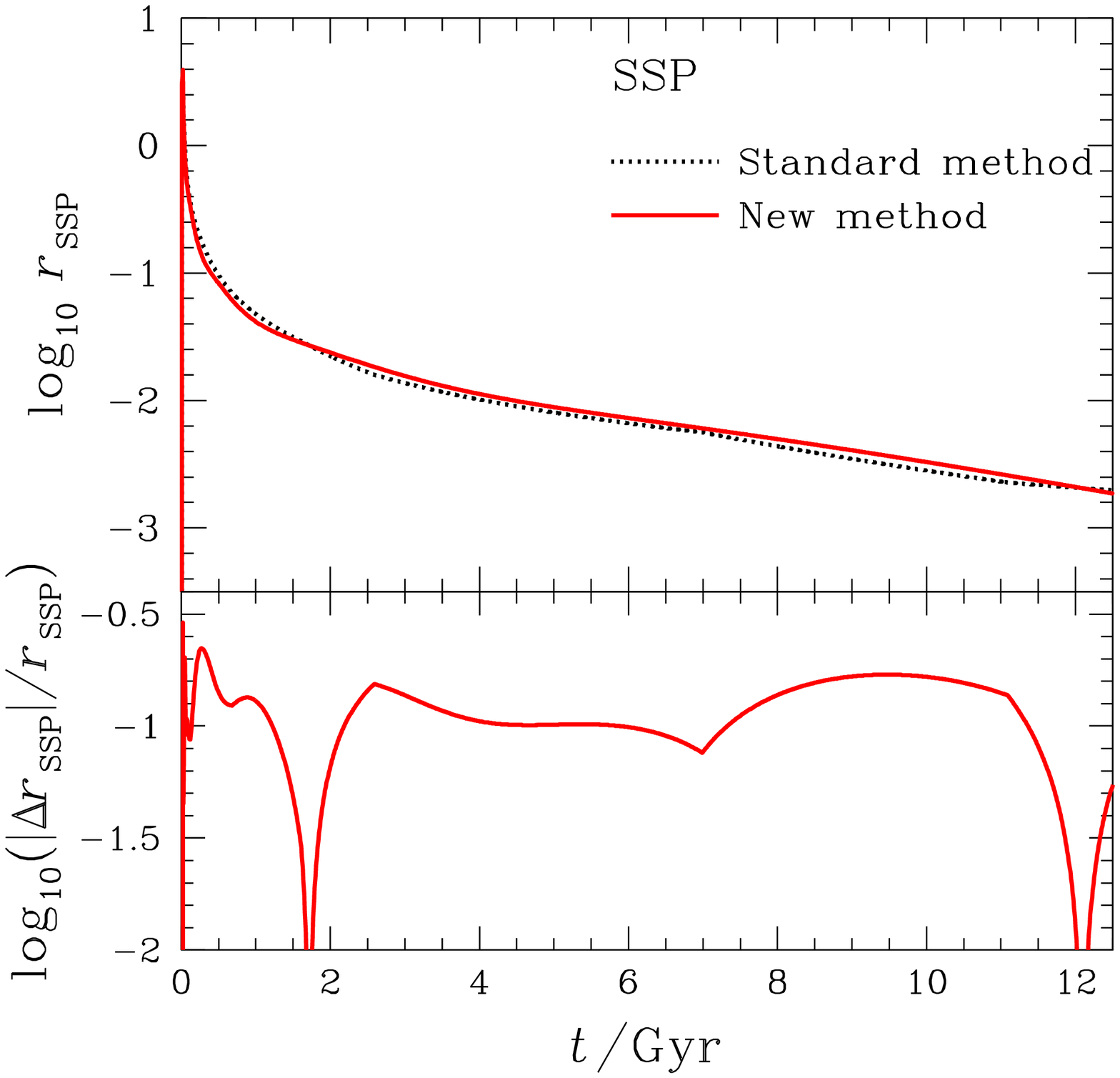,height=7.2cm,width=7.2cm}
\epsfig{file=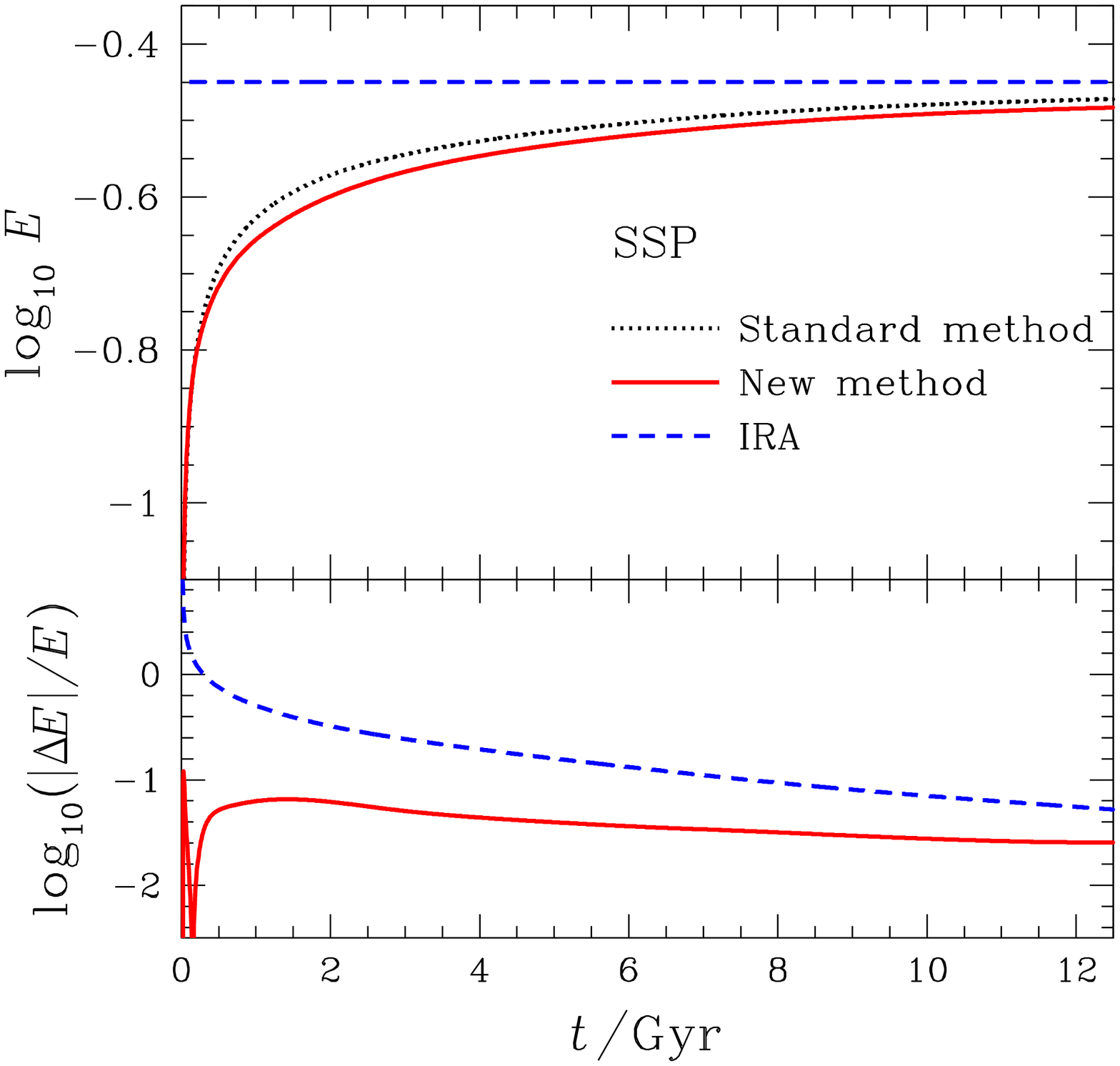,height=7.2cm,width=7.2cm}
\caption{Left panels: exact specific mass return rate $\rssp$ as given
 by eq. (2) (dotted line), along with its fit from eq.~(11) for the
 parameters reported in Table 1 (red solid line). The relative error
 between the two rates is shown in the bottom panel. Right panels:
 the cumulative ejected mass computed from the integral of the the
 exact rate (dotted) and of the fit (solid).  The dashed line
 represents the same quantities calculated assuming the IRA.  In the
 bottom-right panel, the relative errors with respect to the exact
 rate are shown, as computed by means of the new method (solid line)
 and in the case of the IRA.}
\label{mrr_ssp}
\end{figure*}

In Fig.~1 (top panel, short dashed lines) we also show the separate
contribution of the six fit components in eq. (11), whose
parameters $\ai$ and $\bi$ are reported in Tab. 1, together with their sum (long
dashed line). As can be seen from Tab. 1, the best fit parameters obey
\begin{equation}
\sum_{i=1}^6\frac{\ai}{\bi^2}\simeq 0.328
\end{equation}
i.e. the fit in eq. (11) conserves the total mass almost perfectly.
The plot in terms of the logarithm of time allows to
appreciate how well the fit reproduces the exact $\rssp$, with the
exception of the few Myrs after the SSP formation. This is due to the
fact that the exact $\rssp$ presents a discontinuity at the time
corresponding to the lifetime of the highest mass stars (4 Myr for a
$40\Msun$ star): in any case, the functions $t e^{-\beta t}$ allow to
reproduce the initial rise better than pure exponentials, and even
better results would be obtained by using functions $t^n e^{-\beta t}$
with larger values of $n$. Here, for presentation purposes, we
restrict to $n=1$ as a compromise between accuracy and algebraic
simplicity.

The overall impact of the initial discontinuity on the resulting mass
return rate is negligible, as can be seen from Fig. 2.  In the
left-hand panels we show again the exact rate $\rssp$ obtained by
evaluation of eq. (2) (top panel, dotted line) compared to the rate
obtained by summing the six fit components (red solid line), and the
relative error between the fit function and the exact rate $\rssp$ as
a function of time (bottom panel).  In the right-hand panels we show
the cumulative returned mass $E(t)$ and the associated relative error
for the exact $\rssp$ (dotted line), for the new method (red solid
line), and in the case of IRA (blu dashed line), respectively. Note
how the relative error of the new method is significantly smaller than
the one obtained with the IRA, especially at early times.
\begin{table}
\vspace{0cm}
\begin{flushleft}
  \caption[Parameters of the six functions $\ai\, t \, e^{-\bi t}$
  used for the fit of the $\rssp$ considered in Section 3.1.]
  {Parameters of the six functions $\ai\, t \, e^{-\bi t}$ used for
    the fit of the $\rssp$ considered in Section 3.1.}
\begin{tabular}{l|l}
\noalign{\smallskip}
\hline
\hline
\noalign{\smallskip}
  $\ai$  (Gyr$^{-2}$)         &  $\bi$ (Gyr$^{-1}$) \\
\noalign{\smallskip}                                                                                                                                         
\noalign{\smallskip}                                                                                                                                         
\hline                                                                                                                                                       
\noalign{\smallskip}        
  $7.7624\times 10^{-3}$ &$3.1623\times 10^{-1}$\\
  $9.5499\times 10^{-2}$ &$1.2589~~~~~~~~~~~~~$\\
  $1.3183~~~~~~~~~~~~~$ &$5.0119~~~~~~~~~~~~~$\\
  $21.380~~~~~~~~~~~~~$ &$19.953~~~~~~~~~~~~~$\\
  $2.9512\times 10^{2}$  &$79.433~~~~~~~~~~~~~$\\
  $3.7153\times 10^{3}$  &$3.1623\times 10^{2}$\\
\hline
\hline
\end{tabular}
\label{tab_exp}
\end{flushleft}
\end{table}

\subsection{Multiple stellar population}
\label{multiple}

\begin{figure*}
\centering
\epsfig{file=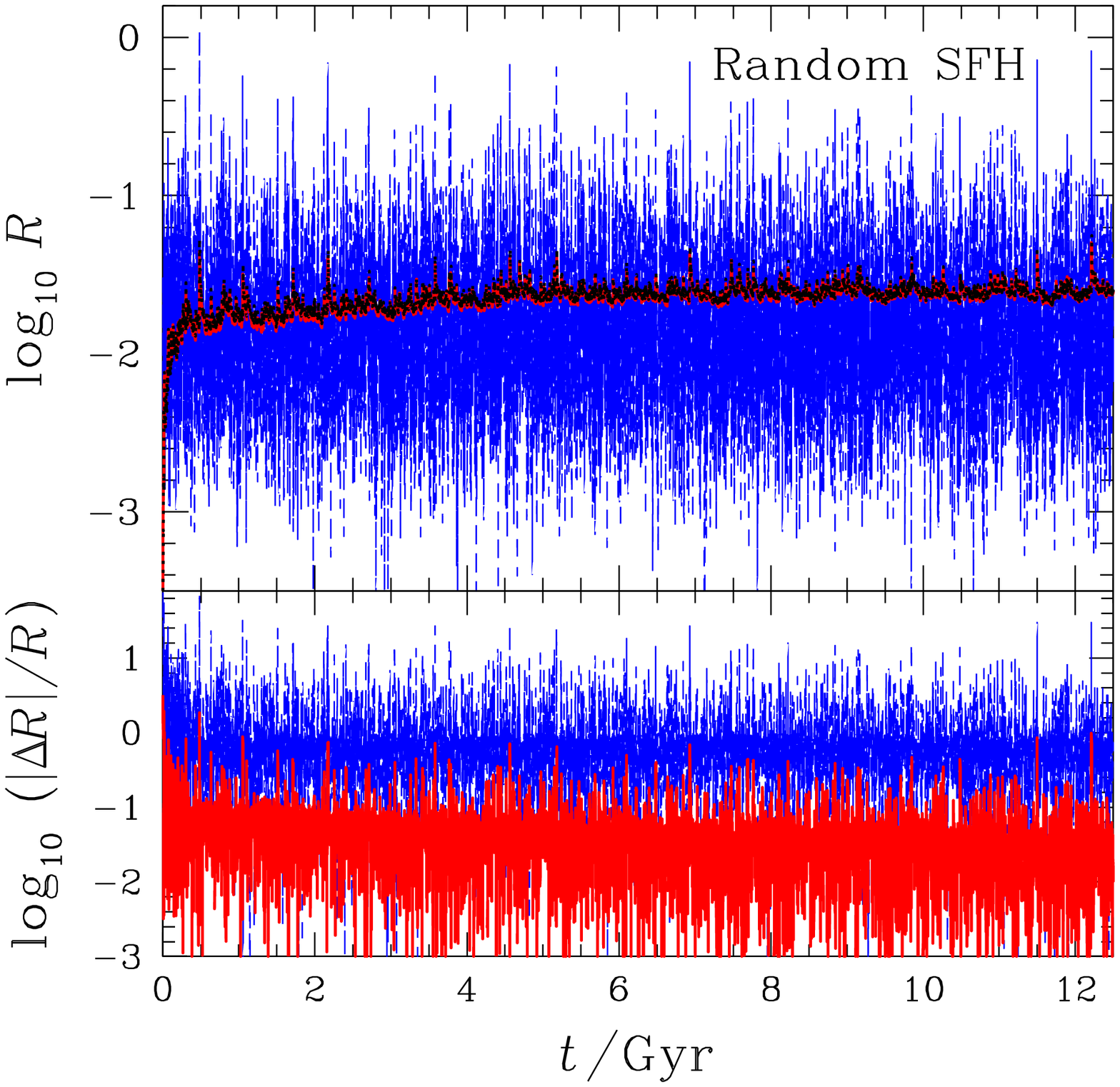,height=7.2cm,width=7.2cm}
\epsfig{file=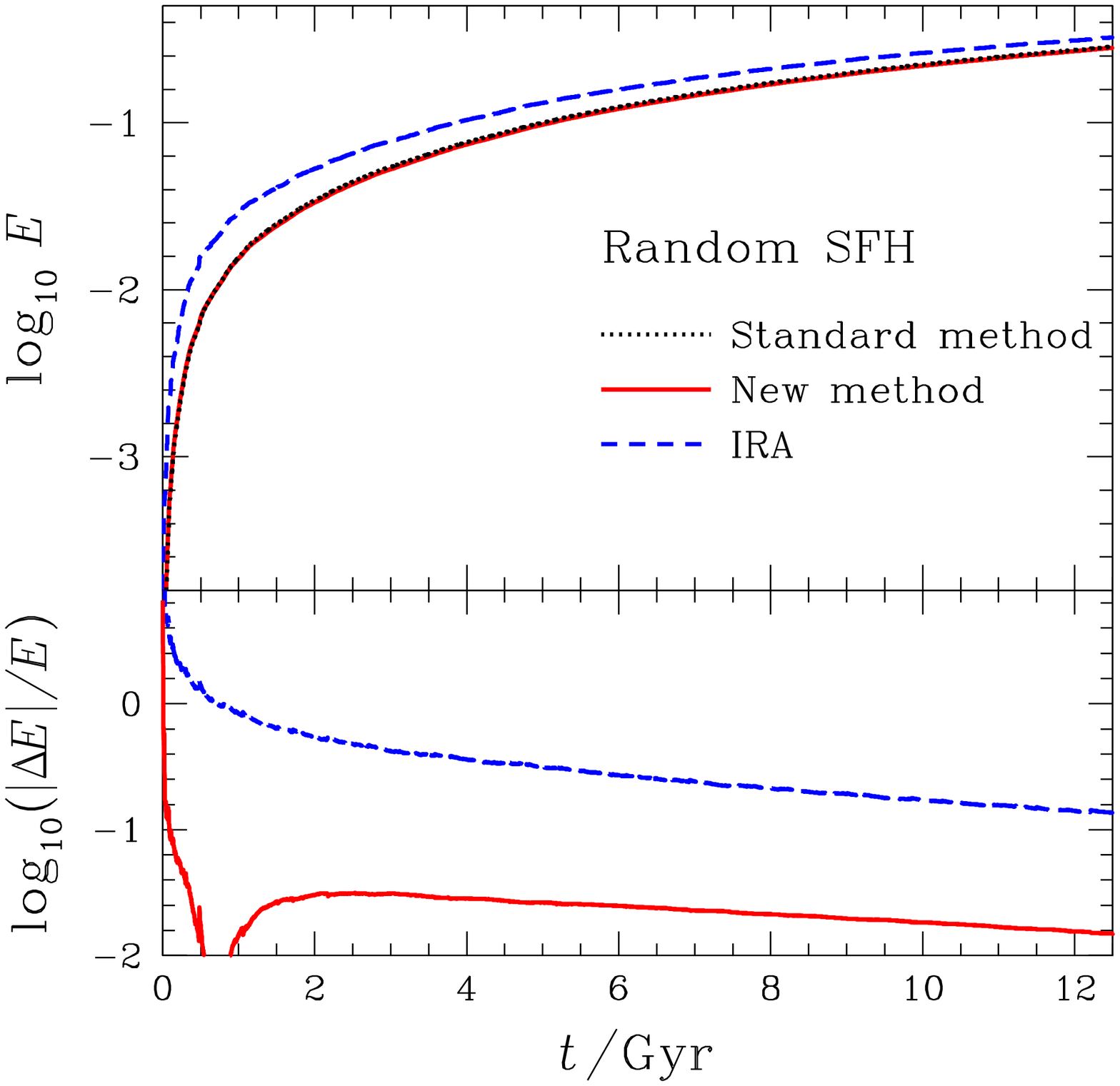,height=7.2cm,width=7.2cm}
\caption{Left panels: time evolution of the mass return rate $R$ of a
  multiple stellar population originated by a SFH consisting of
  several bursts of random amplitude, as described in Sect. 3.2.  The
  black solid line is the rate computed with the standard method, the
  red line with the new method, and the blue line by using the
  IRA. This latter line also visualizes the adopted SFH, because from
  eq. (4) $R_{\rm IRA}(t)\propto \psi (t)$.  The relative error of the
  mass return rate is shown in the bottom panel for the new method
  (red) and assuming the IRA (blue). Right panels: the cumulative mass
  return $E$ and the relative errors are shown, with the same line
  styles as in the left panels. The quantities $R$ and $E$ are
  normalized to the total stellar mass formed over the simulation, and
  the time-step is $\Delta t = 1.4$ Myr.}
\label{rate1.4}
\end{figure*}

We now move to consider the case of the mass return rate of a multiple
stellar population originated by a complex SFH.

In order to test the performance (in terms of computational speed and
accuracy) of the new method, we consider an artificial case,
characterized by a large number of star-formation bursts of random
amplitude.  This case is relevant since it mimics the SFHs commonly
encountered in SAMs, and which are typical of systems undergoing a
large number of mergers and interactions, such as the ones resulting
from the intricate merging trees of giant galaxies (Somerville \&
Kolatt 1999; Lanzoni et al. 2000; Calura \& Menci 2009; Yates et
al. 2013), or star formation induced by AGN activity in the inner
regions of the host systems (Ciotti \& Ostriker 2007). In practice, we
adopt a SFH $\psi(t)$ in which, at each time-step, $\log_{10}\psi$ is
extracted randomly from a normal distribution with standard deviation
$0.5$.  In the following experiment, the time-step is kept constant,
$\Delta t = 1.4$ Myr, suited to resolve the time contribution to the
mass return rate of the most massive stars at early times (in
hydrodynamical simulations considerably shorter time-steps are common,
down to $10^3-10^4$ yr or less).

The behaviour of $\psi (t)$ is visualized by the blue dashed line in
Fig.~3 (top-left panel), representing the mass return rate in the IRA
(which is proportional to $\psi$, see eq. 4). The resulting mass
return rate is also shown for the the standard method (black solid
line) and with the new method (red line): the relative errors with
respect to the exact values of the mass return rate are shown in the
bottom-left panel. The most striking feature is the large scatter in
$R_{\rm IRA}$ (more than two orders of magnitude) compared to the
exact rate, which instead is almost perfectly matched by the rate
obtained with the new method. The relative errors represented in the
bottom-left panel quantify the performance of IRA and of the new method: in
general, the peaks in the relative errors of the IRA are related with
peaks in the instantaneous SFH $\psi(t)$. Such peaks are also present
in the mass return computed with the new method, as a consequence of
the initial discontinuity following each burst, as described
above. However, the amplitude of these error is significantly smaller
than in the IRA case, as apparent.

The cumulative mass return for the multi-burst SFH is shown in the
right panels of Fig.~3, where the quantity $E(t)$ is normalized to the
stellar mass formed over the entire simulation. The top-right panel shows
this quantity for the exact case, for the IRA case, and for the new
method, while the bottom-right panel shows the relative error between the
IRA and the exact case, and between the new method and the exact case.
Note the remarkable accuracy of the new method in the multiple
population case, actually even better than in the case of a SSP. This
is not surprising, as for a multiple stellar population the IRA is, in
practice, always in the most critical regime (just after a burst), and
its error is of a magnitude similar to that presented at early times
in Fig. 2 (bottom right panel).

\subsection{Computational advantages of the new method}
\label{cpu}

As already discussed, the IRA presents considerable advantages in
terms of computational time and does not require any storage regarding
ages and metallicities of composite stellar populations. Essentially,
at any time, the mass return rate is calculated directly from the
physical properties of the system at that particular time,
i.e. directly from the instantaneous value of the star formation rate
(see eq. 4). The major problem with the IRA is that it breaks down at short times after a starburst event, when the very short characteristic
heating and cooling times would lead to large overestimates of the
mass return rate, that in turn can seriously affect the numerical
computations by altering the available mass budget.

The method presented in this paper offers the possibility to compute
the mass return rate for arbitrarily complicated SFHs, at a
computational cost considerably lower than that required when the
exact method is used, yet maintaining the full accuracy. The
computational gain of the new method with respect to the standard
method can be estimated by considering the number of needed
operations. The computational time of the exact calculation scales
with the total number of time-steps $\Nstep$ as $T_{\rm CPU,standard}
\propto \Nstep(\Nstep+1)/2$, since $j$ operations are needed to
compute the mass return rate at the $j$-th time-step. With the new
method, $T_{\rm CPU,new} \propto (n+1)k\Nstep$, because at each time
the $n+1$ integrals over the last time step $\Jz,\ldots,J^{(n)}$ must
be computed for each of the $k$ functions used in the fit. Therefore,
$T_{\rm CPU,new} /T_{\rm CPU,standard}\sim 2(n+1)k/\Nstep$: when the
other parameters are fixed, the computational gain of the new method
over the standard method increases linearly with $\Nstep$, i.e. it
increases for decreasing $\Delta t$ (at fixed total time span). 
Note that the CPU time of the new method scales
with $\Nstep$ as in the IRA.

The actual computational gain of the new method obtained in numerical
experiments similar to that described in Sect. 3.2, where different
$\Delta t$ are adopted, is shown in Fig.~4, plotting the CPU time (in
arbitrary units) required to evaluate the mass return for the exact
method (dotted line), the new method (red line), and the IRA (dashed
line), as a function of the total number of time-steps $\Nstep$.  For
our reference case with $\Delta t=1.4$ Myr, we have $\Nstep=8929$,
$k=6$, $n=1$, so we expect $T_{\rm CPU,new} /T_{\rm
  CPU,standard}\simeq 2.7\times 10^{-3}$. In fact, the experiment
shows that the CPU time for the new method is a factor $\approx 500$
shorter than for the standard method, in agreement with the analytic
estimate.  From Fig.~4 it is also apparent that the actual scaling
with $\Nstep$ nicely follows the expected scaling ($T_{\rm
  CPU,new}\propto T_{\rm CPU,IRA}\propto\Nstep$, $T_{\rm
  CPU,standard}\propto \Nstep^2$). We remark again that in
hydrodynamical simulations the actual $\Delta t$ can easily be one or
two orders of magnitude shorter than 1 Myr, so that the gain of the
new method will increase accordingly; moreover, this further gain will
also be multiplied by the number of grid points.

\begin{figure}
\epsfig{file=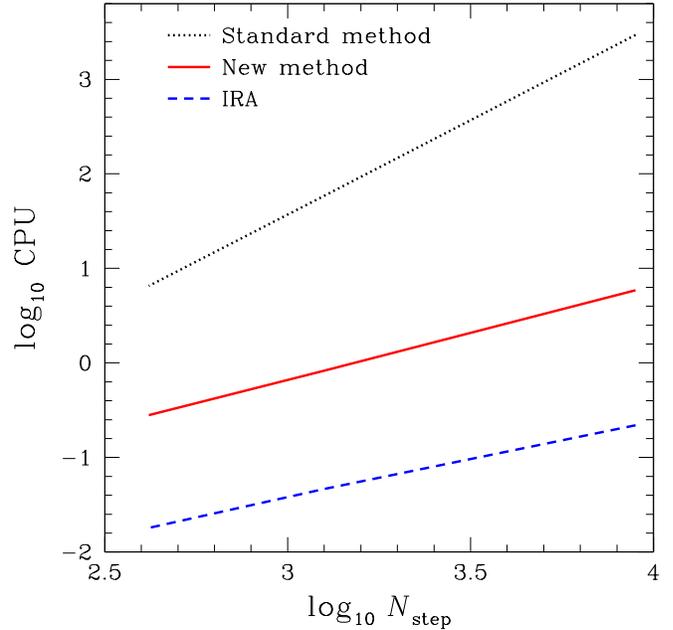,height=9.2cm,width=9.2cm}
\caption{CPU time (in arbitrary units) required by the computation of
  the mass return rate by means of the standard method (dotted line),
  the new method (solid line) and the IRA (dashed line), as a function
  of the total number of time-steps $\Nstep$.  The case discussed in
  Sect. 3.2 corresponds to $\log_{10}\Nstep \simeq 3.95$.}
\label{fig_cpu}
\end{figure}

\section{Conclusions} 

The mass returned to the ambient medium by evolving stellar
populations represents a significant fraction of their initial mass
(20\% to 30\% for realistic IMFs), and thus being a non negligible
contribution to the evolution of stellar systems. For example, in
elliptical galaxies the total mass of the returned gas is almost two
orders of magnitude larger than the observed masses of central SMBHs,
requiring important AGN feedback effects to prevent accretion in
absence of other mechanisms able to eject the metal rich gas from the
galaxies (e.g., Ciotti \& Ostriker 2001).

In numerical problems dealing with the evolution of stellar
populations and their mass return, when the characteristic times are
longer than the stellar evolutionary times, the IRA is a useful
approximation of the convolution integral describing mass return,
allowing for fast and accurate computations. However, in
hydrodynamical studies of star formation, of the interplay between
central starbursts and AGN feedback in the coevolution of galaxies and
their central SMBHs, and of the origin of multiple stellar populations
in globular clusters, the dominance of short characteristic times
($\simeq 10^3$ yr of even less), coupled with thousands (or more)
spatial grid points, requires the detailed evaluation of the
convolution integral. This makes the simulations excessively demanding
in terms of computer memory and computational time, due to the need of
storing the whole SFH at each grid point and to the number of
evaluations to be performed increasing quadratically with the number
of time steps.

In order to overcome these problems we developed a fast and accurate
method to evaluate the mass return rate from stellar populations with
arbitrarily complicated SFHs. The method presents great computational
advantages over the direct evaluation of the convolution integral (the
computation time scales linearly with the number of time steps), yet
its accuracy is much better than that achievable with the IRA,
especially in the case of multiple populations, when the clock of the
mass return rate is reset at each star formation event. The new method
can be easily implemented in semi-analytical models and in grid-based
hydrodynamical simulations, and it requires to fit the mass return
rate from the adopted SSP by means of simple functions, solutions of
linear ODEs with constant coefficients.  We have shown how these
functions allow to evaluate, at any time-step, the mass return rate
directly from its value at the previous time step, plus a contribution
depending only on the star formation over the last time step. The
general mathematical framework is presented in the Appendix. In this
paper, as a specific application, we have shown that the use of a
linear combination of the functions $t \, e^{-\beta t}$ is enough to
achieve very accurate results: with arbitrarily complicated SFHs, the
computed mass return rate from the resulting multiple stellar
population deviates by amounts of a few percent from the exact value,
and this avoiding the storage of the SFH itself.

We conclude by noticing that the presented method, applying in general
to the treatment of time convolution integrals, can be also used in
problems different from that discussed in this paper, for example to
study the return of single chemical elements from multiple stellar
populations (as in starbursts), or to describe the time evolution of
the SNIa rate in star forming events.

\section*{Acknowledgments}

We are grateful to Laura Greggio, Francesca Matteucci, Alvio Renzini
and Monica Tosi for useful comments. Financial support from PRIN MIUR
2010-2011, project ``The Chemical and Dynamical Evolution of the Milky
Way and Local Group Galaxies'', prot. 2010LY5N2T is acknowledged.

\appendix

\section{The general ODE for the mass return rate}

For ease of notation we rewrite eq. (3) as
\begin{equation}
Y(t)=\int_{-\infty}^t\psi(\tau)y(t-\tau)\,d\tau,
\end{equation}
where the meaning of $Y$ and $y$ is obvious, and the lower limit of
integration allows full generality in the treatment. Let assume that
$y$ is a solution of a homogeneous, constant-coefficients linear ODE of order $n$, i.e.,
\begin{equation}
\sum_{i=0}^n\ci\yi =0,
\end{equation}
where $\ci$ are constants, $\yi \equiv d^iy/dt^i$, and $y^{(0)}\equiv
y$. Taking for granted that all the technical requirements of
convergence and regularity are satisfied, mathematical induction
proves that the time derivatives of the function $Y$ can be written as
\begin{equation}
Y^{(i)}(t)=\sum_{j=0}^{i-1}\psi^{(i-j-1)}(t) y^{(j)}(0) + \int_{-\infty}^t\psi(\tau)\yi(t-\tau)\,d\tau,
\end{equation}
for $i=1,2,...,n$.  After multiplication of eq. (A3) by $\ci$ and
summation over $i$, from eq. (A2) it follows that $Y$ also obeys the
generally non-homogeneous constant-coefficients linear ODE of order
$n$
\begin{equation}
\sum_{i=0}^n\ci Y^{(i)}=\sum_{i=0}^n\sum_{j=0}^{i-1}\ci\psi^{(i-j-1)}(t) y^{(j)}(0).
\end{equation}
Note that the assumption of constant coefficients is important here,
because the vanishing of the resulting linear combination of integrals
at the r.h.s. of eq. (A3) rests on two facts: 1) the $\ci$ can be
carried under sign of integration (this would be true also for
time-dependent coefficients $\ci$), and 2) if $y(t)$ is a solution of eq. (A2),
then also $y(t-\tau)$ is a solution (this in general would not be true
for time-dependent $\ci$).  Therefore, if the kernel
$\rssp$ is solution of a constant-coefficients linear ODE of order
$n$, also the mass return rate $R$ can be represented as a solution of
constant-coefficients linear ODE of order $n$, and so $n$ integration
constants are required for its complete determination.

Suppose now to consider the following generalization of eqs. (5) and (11)
\begin{equation}
y(t)=\sum_{i=1}^k\sum_{j=0}^{\nei} \aij t^je^{-\bi t},
\end{equation}
where $\aij=0$ for $j>\nei$, $\aij\neq 0$ for $j=\nei$, and for
$j<\nei$ the coefficients $\aij$ may be zero or not; without loss of
generality we assume that all the $\bi$ are different. Note that in
eq. (A5) we consider the possibility that more than one power term
$t^j$ is associated with a given $\bi$, and that the maximum values
$\nei$ differ for different $\bi$. In this formalism eq. (5) is
obtained for $\nei=0$ and $\alpha_{i0}=\ai$ ($ i=1,...,k$), while
eq. (11) for $\nei=1$, $\alpha_{i0}=0$, and $\alpha_{i1}=\ai$
($i=1,...,k$).

We now construct the ODE for $y$ in eq. (A5), so that also the ODE for
the mass return rate $R(t)$ can be explicitly obtained in this general
case accordingly to eq. (A4). From the theory of ODEs (e.g., Ince
1956) it is known that the function $t^n e^{-\beta t}$ is solution of
a constant-coefficients linear ODE, where the root $-\beta$ of the
associated characteristic polynomial has (at least) algebraic
multiplicity $n+1$. Therefore, the minimum characteristic polynomial of
the ODE admitting eq. (A5) as solution is
\begin{equation}
\prod\limits_{i=1}^k (\lambda+\bi)^{\nei+1}=0,
\end{equation}
and the corresponding ODE of order $\sum_{i=1}^k(\nei+1)$ is
obtained by the substitution $\lambda=d/dt$.

A final comment is in order. Due to the formal similarity of eq. (A1)
with the convolution integral appearing in the theory of Green
functions, one may ask whether $y(t)\theta(t)$ (where $\theta$ is the
Heaviside step function) is the Green function, vanishing for $t<0$,
of the ODE (A2). The answer is negative, as $y(t)$ in general lacks
the needed regularity at $t=0$ (e.g., Bender \& Orszag 1978). In fact,
from the request of vanishing for $t<0$, it follows that the function
$y(t)\theta (t)$ is a genuine Green function if and only if the time
derivative of order $\sum_{i=1}^k(\nei+1) -1$ of $y(t)$ evaluated at
$t=0$ presents a finite jump, while all the remaining derivatives of
$y(t)$ up to the order $\sum_{i=1}^k(\nei+1) -2$ evaluated at $t=0$
vanish. These requests impose constraints on the values
$\aij$. However, in our method the $\aij$ are given as best-fit
parameters of the kernel function (in practice, by the request of
reproducing $\rssp$), so that $y(t)$ in general does not satisfy the
regularity conditions at $t=0$. However, it is easy to show that in
the case of a single-component fit ($k=1$), the function $t^n
e^{-\beta t}\theta (t)$ with $n\geq 0$ {\it is} the Green function for
the corresponding linear ODE of order $n+1$.

\label{lastpage}

\end{document}